\documentclass[a4paper]{jpconf}
\usepackage{graphicx}
\begin{document}
\title{Background model of the ANAIS-112 dark matter experiment}

\author{J~Amaré$^{1,2}$, S~Cebrián$^{1,2}$, D~Cintas$^{1,2}$, I~Coarasa$^{1,2}$, E~García$^{1,2}$, M~Martínez$^{1,2,3}$, M~A~Oliván$^{1,2,4}$, Y~Ortigoza$^{1,2,5}$, A~Ortiz de Solórzano$^{1,2}$, J~Puimedón$^{1,2}$, A~Salinas$^{1,2}$, M~L~Sarsa$^{1,2}$ and P~Villar$^{1}$}

\address{$^{1}$ Centro de Astropartículas y Física de Altas Energías (CAPA), Universidad de Zaragoza, Pedro Cerbuna 12, 50009 Zaragoza, Spain \\
$^{2}$ Laboratorio Subterráneo de Canfranc, Paseo de los Ayerbe s.n.,
22880 Canfranc Estación, Huesca, Spain \\
$^{3}$ Fundación ARAID, Avenida de Ranillas 1D, 50018 Zaragoza, Spain \\
$^{4}$ Fundación CIRCE, Avenida de Ranillas 3D, 50018 Zaragoza, Spain\\
$^{5}$ Escuela Universitaria Politécnica de La Almunia de Doña Godina (EUPLA), Universidad de Zaragoza, Calle Mayor 5, La Almunia de Doña Godina, 50100 Zaragoza, Spain}

\ead{scebrian@unizar.es}

\begin{abstract}
The ANAIS (Annual modulation with NaI(Tl) Scintillators) experiment aims at the confirmation or refutation of the DAMA/LIBRA positive annual modulation signal in the low energy detection rate. ANAIS-112, consisting of nine 12.5~kg NaI(Tl) modules, is taking data since August, 2017 at the Canfranc Underground Laboratory (LSC) in Spain. Results from the analysis of three years of data are compatible with the absence of modulation. The background model developed for all nine ANAIS-112 detectors was established from commissioning data and non-blinded events in the first year of data taking. Now, background characterization is being improved profiting from the larger accumulated exposure available. Here, the background model is described and comparisons of model and measurements for energy spectra and counting rate time evolution for three-year exposure (considering different analysis conditions) are presented. 
\end{abstract}

\section{Introduction}

The observation by the DAMA/LIBRA experiment of an annual modulation signal in the counting rate, compatible with expectations from galactic dark matter particles due to Earth’s movement around the Sun, has intrigued the community for more than twenty years \cite{damai,dama}. The ANAIS-112 experiment, with a target of 112.5~kg of NaI(Tl) \cite{epjc2019perf}, is running smoothly at the Canfranc Underground Laboratory since 2017 aiming to test this observation using the same detection technique and target. A 3$\sigma$ sensitivity to explore the DAMA/LIBRA result for five-year operation is expected \cite{epjc2019sen}. After the first annual modulation analysis corresponding to 1.5~years \cite{prl}, results from three years of data have been presented in 2021 \cite{prd}. Under the hypothesis of modulation, the deduced amplitudes from best fits are in all cases compatible with zero for the two energy regions considered at 2–6 and 1–6 keV\footnote{Electron equivalent energy is given throughout the paper.}.

A full description of the ANAIS-112 set-up and its performance can be found in \cite{epjc2019perf}. The experiment consists of an array of 3$\times$3 NaI(Tl) scintillators (named D0 to D8) manufactured by the Alpha Spectra Inc. company in Colorado (US); each crystal, with a mass of 12.5~kg, is coupled to two Hamamatsu photomultipliers (PMTs) and enclosured in a copper vessel. The shielding of ANAIS-112 is made of 10~cm of archaeological lead, 20 cm of low-activity lead, a box filled with radon-free N$_2$ gas to avoid radon intrusion, and 40~cm of neutron moderator. Sixteen plastic scintillators covering  the set-up act as an active muon veto. The LSC facilities are placed at a depth of 2450~m.w.e. The coincidence (in a 200~ns window) of the two PMT signals from a module provides the acquisition trigger to digitize all PMT electric pulses. The analysis threshold is set at 1~keV thanks to an outstanding light collection of $\sim$15 photoelectrons per keV for all detector units and an effective rejection of non-scintillation events.

The background of ANAIS-112 detectors is being analyzed since the beginning of the operation at LSC of the first modules D0 and D1 \cite{epjc2016} and models for all detectors were developed from non-blinded events (corresponding to multiple-hits and to single-hit events releasing energy above 6~keV) after the first year of data taking \cite{epjc2019}. They are based on the Geant4 simulation, including a detailed description of detectors and shielding, of the main background sources quantified applying different techniques. Simulated energy spectra have been cross-checked against measured data at different conditions and energy ranges. Now, the background characterization is being improved thanks to the exposure accumulated over several years, which allows for time-dependent analysis.

\section{Background sources}
The radioactivity of external components like PMTs and copper enclosures was measured with HPGe detectors at LSC and included in the models. But the dominant background contribution is the intrinsic activity of NaI(Tl) crystals, which has been directly assessed for each ANAIS-112 detector in several set-ups. $^{40}$K activity was quantified by identifying coincidences among different modules \cite{ijmpaK}, ranging the measured values from (1.33$\pm$0.04)~mBq/kg to (0.54$\pm$0.04)~mBq/kg. The content of $^{232}$Th and $^{238}$U was determined from the measured alpha rate following Pulse Shape Analysis and the study of BiPo sequences, being the activity of the chain isotopes at the level of a few $\mu$Bq/kg except for $^{210}$Pb, which was found to be out of equilibrium with measured activity between (3.15$\pm$0.10)~mBq/kg and (0.7$\pm$0.1)~mBq/kg. Cosmogenic isotopes induced in the crystals when being on surface have also been carefully studied \cite{jcap,ijmpacosmo}, including several short-lived Te and I isotopes which have already decayed and $^{109}$Cd and $^{113}$Sn, producing peaks at the binding energies of K-shell electrons (following Electron Capture); for these two isotopes, the saturation activity assumption firstly considered has been now relaxed according to the observed time evolution of the peaks. The activity of $^{22}$Na was also quantified from the analysis of coincidences from high-energy gamma-rays, being the measured rate at 0.9 keV well reproduced by simulation using the independently quantified activity. An additional background source contributing only in the very low energy region was identified, compatible with an initial tritium activity \cite{tritium} at the level of 0.20~mBq/kg for the first two crystals produced and 0.09~mBq/kg for the others.

\section{Results from three-year exposure}
After the first validation of the background model of the ANAIS-112 detectors carried out using the first year of data \cite{epjc2019}, new comparisons between simulation and data collected from 2017 to 2020 (313.95 kg$\times$y) have been made not only for energy spectra but also for the time evolution of counting rates in different energy regions.

\begin{figure}
\begin{center}
\includegraphics[width=0.48\textwidth]{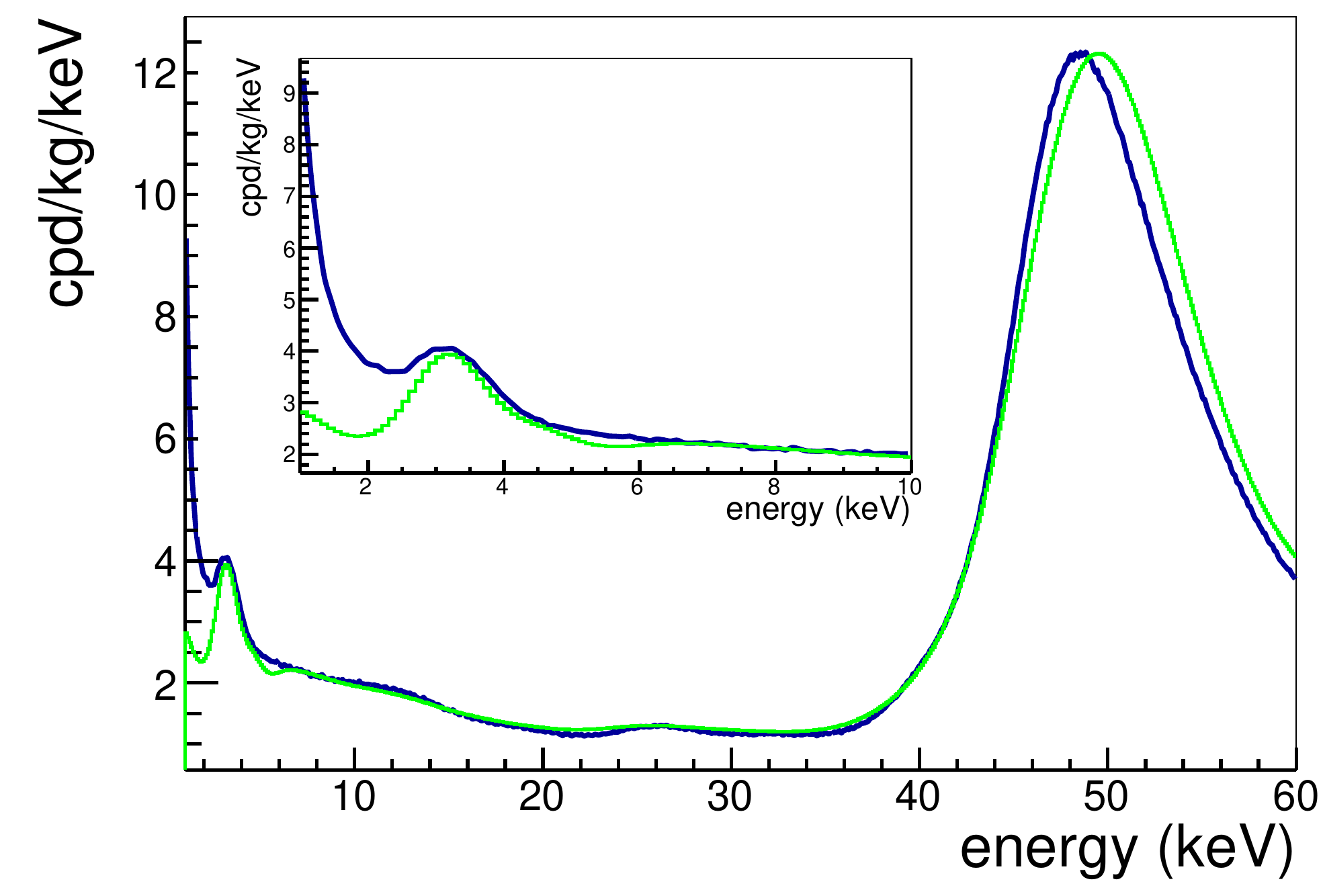}
\includegraphics[width=0.48\textwidth]{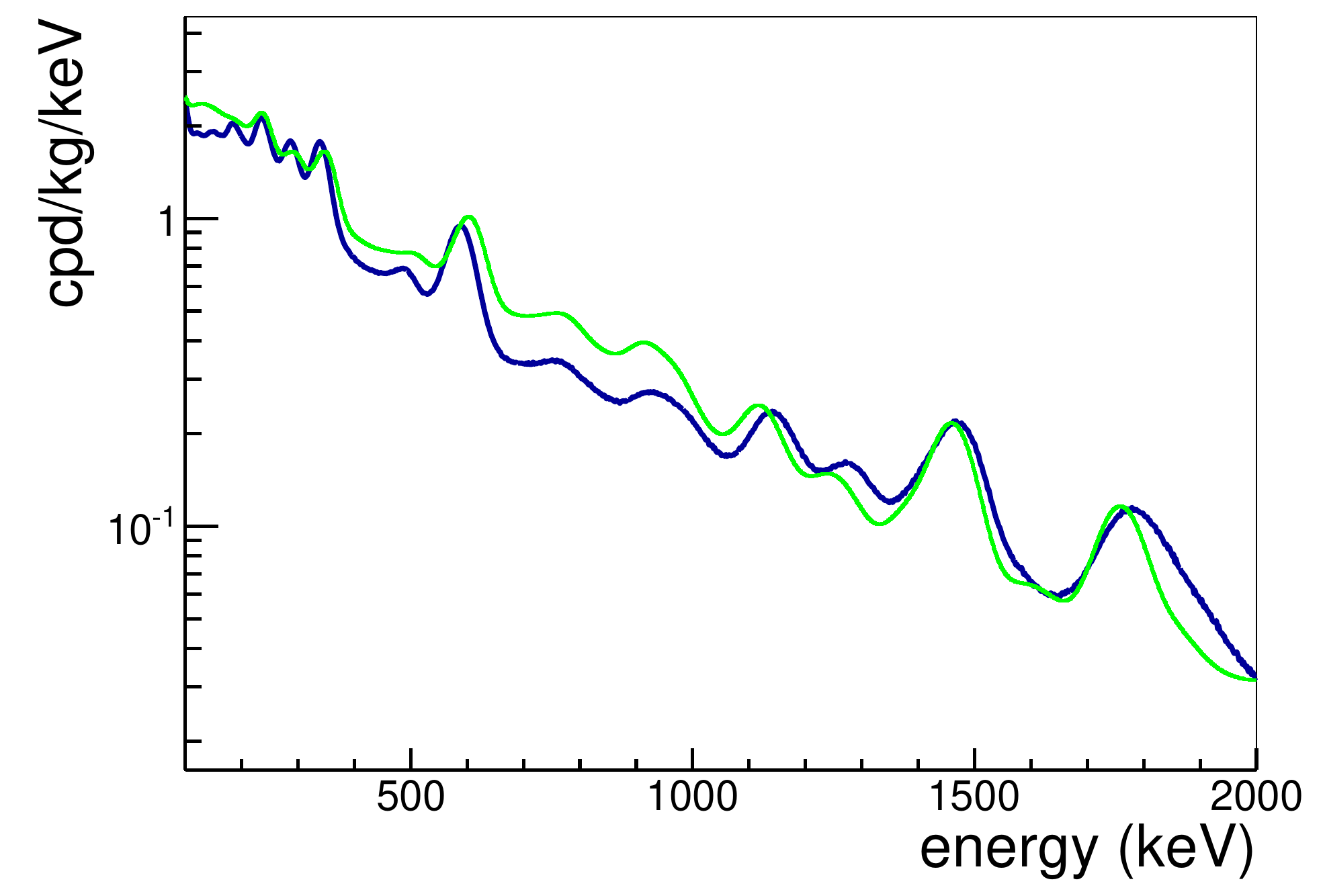}
\end{center}
\caption{\label{spectra} Comparison of the measured spectra at different energy regions in the three-year exposure data of ANAIS-112 (blue line) with the corresponding estimate from the background model (green line).}
\end{figure}

Figure \ref{spectra} compares the ANAIS-112 spectra measured in different energy regions with the model predictions; the low energy spectrum corresponds to events with multiplicity 1 (M1) including the efficiency correction after filtering of non-scintillation events. Similar comparisons have been made for each detector and considering also events with multiplicity 2 (M2), that is, with energy depositions in two detectors. The overall agreement is satisfactory although there are in the data unexplained events below 3~keV, which could be due to an unknown background source not included in the model or to events leaking the filtering procedures. Very promising results are being obtained to partially dispose of these unexplained events when applying machine-learning techniques in the filtering protocols \cite{coarasa}.

The simulation of the different background sources has allowed to quantify the main contributions in the region of interest from 1 to 6 keV, being the most significant ones contaminations in the crystals: the continua from $^{210}$Pb (bulk+surface) and $^{3}$H (with 32.5\% and 26.5\% of the rate, respectively) and peaks from $^{40}$K and $^{22}$Na (with 12.0\% and 2.0\% each).   

\begin{figure}
\begin{center}
\includegraphics[width=0.48\textwidth]{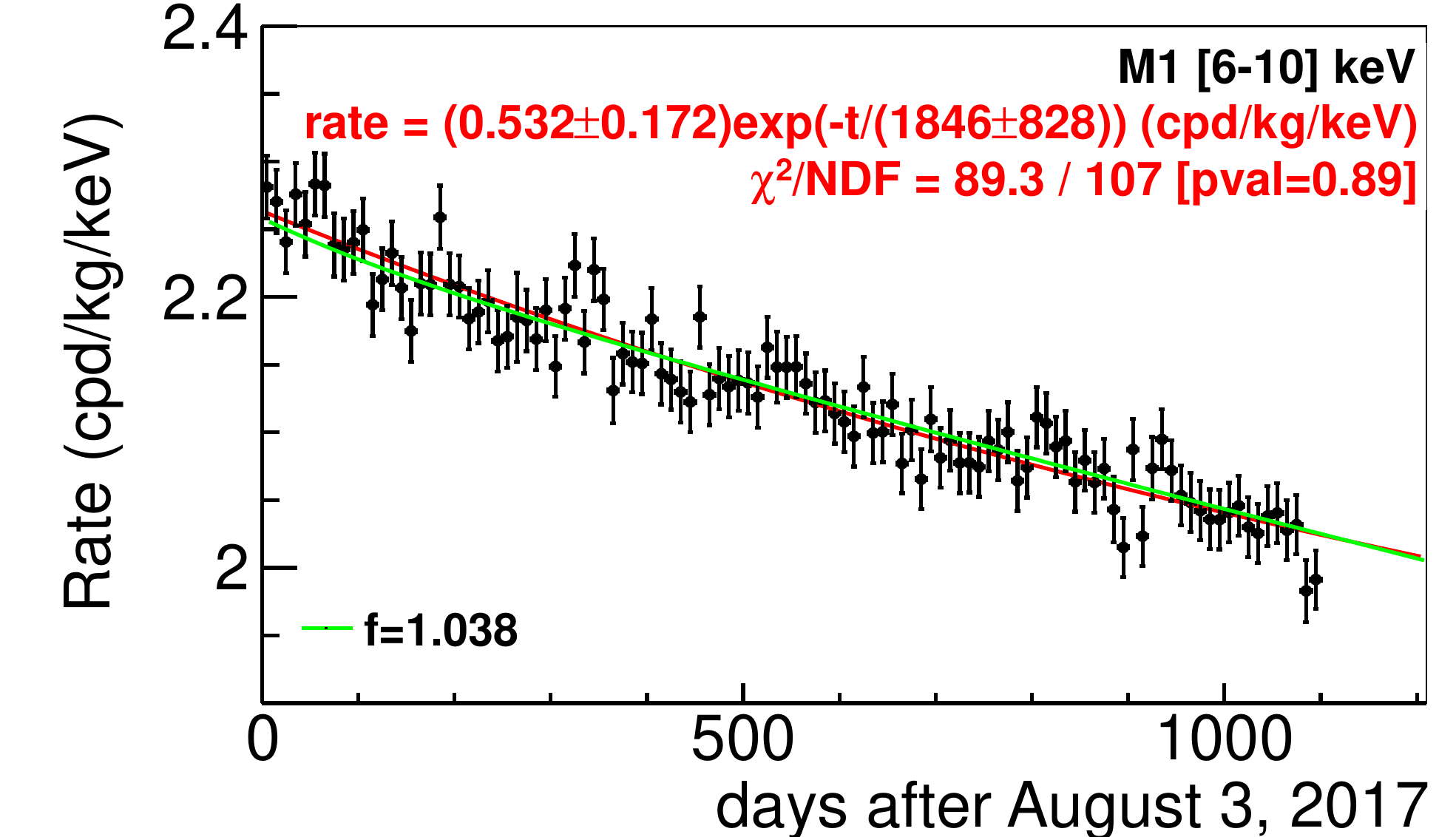}
\includegraphics[width=0.48\textwidth]{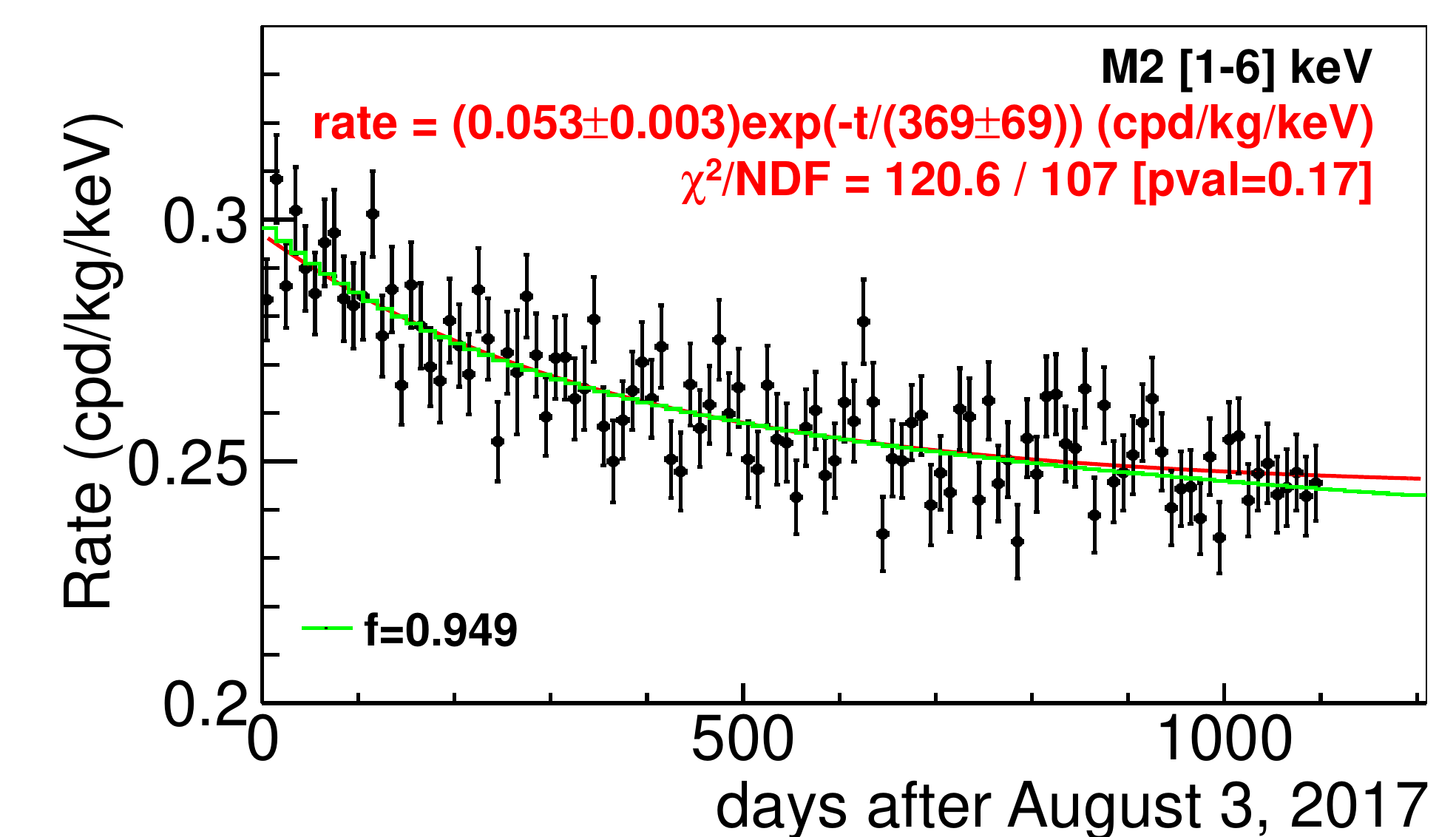}
\end{center}
\caption{\label{rates} Time evolution over the three-year data taking of ANAIS-112 of the measured counting rates (black points), the corresponding exponential fit (red line) and the background model estimate normalized by a factor f (green line), for M1 events at 6-10 keV (left) and M2 events at 1-6 keV (right).}
\end{figure}

\begin{figure}
\includegraphics[width=18pc]{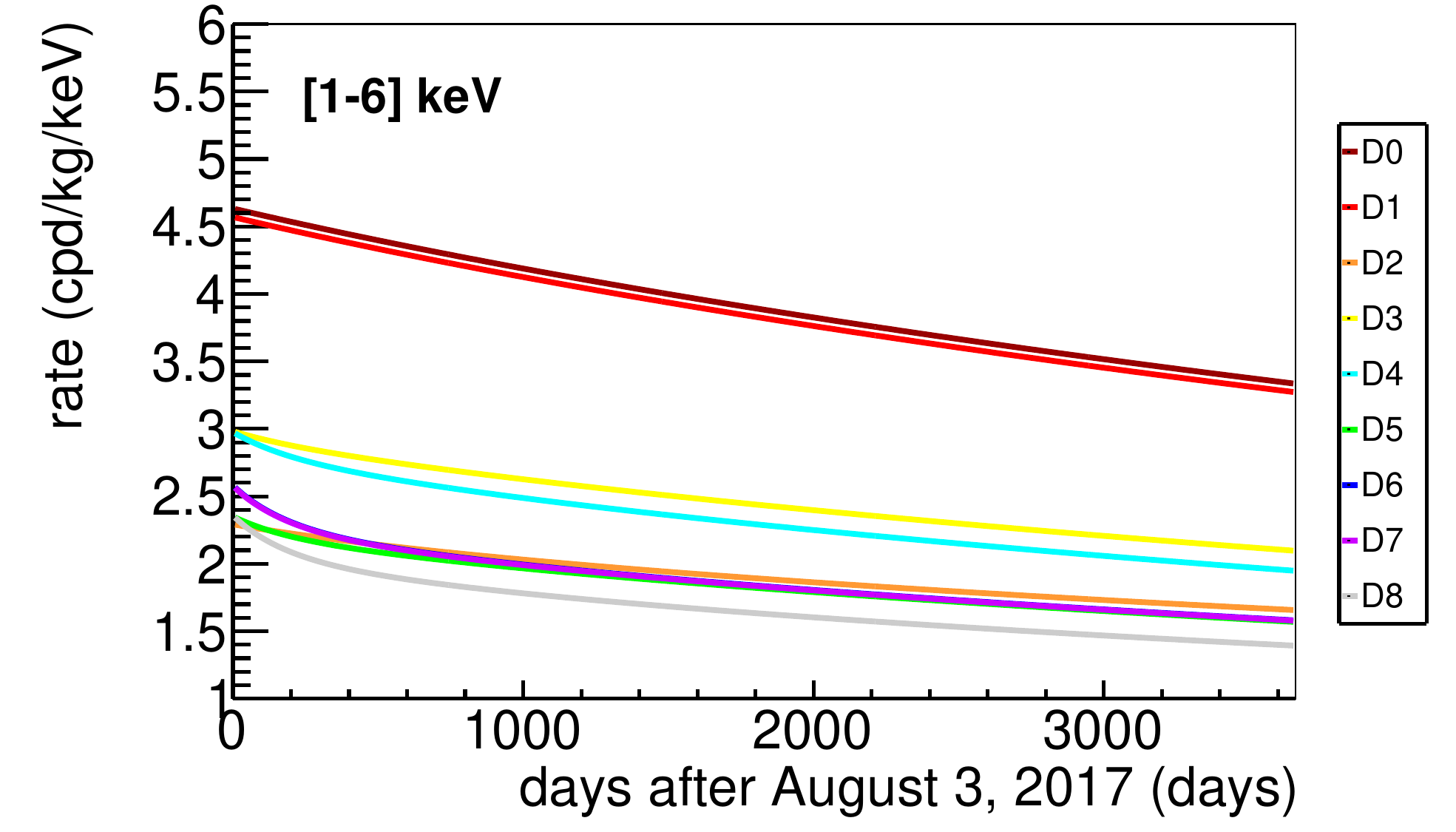}\hspace{2pc}%
\begin{minipage}[b]{18pc}\caption{\label{ratesim} Expected time evolution over ten years of the counting rate of M1 events at 1-6 keV for ANAIS-112 detectors according to their background models.}
\end{minipage}
\end{figure}

Several of the background contributions identified in ANAIS-112 are expected to decrease in time due to the half-life of radioactive isotopes, like $^{210}$Pb (T$_{1/2}$=22.2~y) and cosmogenic $^{3}$H (T$_{1/2}$=12.3~y) and $^{22}$Na (T$_{1/2}$=2.6~y) in the NaI(Tl) crystals. The decrease in counting rates has been observed over the three-year data taking and is accounted for by the background models of the detectors in low energy windows. Figure \ref{rates} presents the evolution in time of the measured rates of M1 events at 6-10 keV and of M2 events at 1-6 keV; the shape of the observed exponential decay is well reproduced by the model, including a normalization factor f, for both event populations even if the ``effective'' lifetimes deduced for each of them are very different. This good agreement gives support to the predicted time evolution of M1 events at 1-6 keV evaluated for each detector, shown in Fig. \ref{ratesim}, which has been considered in the annual modulation analysis corresponding to three-year exposure \cite{prd,martinez}.

\section{Summary}
A good understanding of the background of ANAIS-112 detectors, dominated by NaI(Tl) crystal activity, has been achieved. The energy spectra obtained in different conditions are well reproduced, which has been useful for design and sensitivity predictions. The measured counting rates in the regions 1-6 (2-6)~keV are 3.46 (3.11)~cpd/kg/keV, higher than model prediction by 20 (8.8)\%; the observed excess of events below 3~keV is under study. Additionally, the detector background time evolution observed over the three-year data analyzed is also well described, supporting the model, which has been used in the annual modulation analysis carried out.

\ack{This work has been financially supported by the Spanish Ministerio de Economía y Competitividad and the European Regional Development Fund (MINECO-FEDER) under Grant No. FPA2017-83133-P; the Ministerio de Ciencia e Innovación - Agencia Estatal de Investigación under Grant No. PID2019-104374GB-I00; the Consolider-Ingenio 2010 Programme under Grants No. MultiDark CSD2009-00064 and No. CPAN CSD2007-00042; the LSC Consortium; and the Gobierno de Aragón and the European Social Fund (Group in Nuclear and
Astroparticle Physics and I. Coarasa predoctoral grant). We thank the support of the Spanish Red Consolider MultiDark FPA2017-90566-REDC and acknowledge the use of Servicio General de Apoyo a la Investigación-SAI, Universidad de Zaragoza, and technical support from LSC and GIFNA staff.}

\end{document}